\documentclass[seceq,mbf,epsf,wrapft,amssymb]{ptptex}
\usepackage[dvips]{graphicx}
\usepackage{wrapft}

\title{Improvement of the hot QCD pressure by the minimal 
sensitivity criterion}

\author{
Masahiro {\sc Inui}\footnote{E-mail: inui@sci.osaka-cu.ac.jp}, 
Akira {\sc Ni\'egawa}\footnote{E-mail: niegawa@sci.osaka-cu.ac.jp} 
and Hideaki {\sc Ozaki}\footnote{E-mail: hozaki@sci.osaka-cu.ac.jp}}

\inst{Graduate School of Science, Osaka City University, Osaka 
558-8585, Japan}

\abst{
The principles of minimal sensitivity (PMS) criterion is applied to 
the perturbative free energy density, or pressure, of hot QCD, which 
include the $\sim g_s^6 \ln g_s$ and part of the $\sim g_s^6$ terms. 
Applications are made separately to the short- and long-distance 
parts of the pressure. Comparison with the lattice results, 
at low temperatures, shows that the resultant \lq\lq optimal'' approximants 
are substantially improved when compared to the 
$\overline{\mbox{MS}}$ results. In particular, for the realistic 
case of three quark flavors, the \lq\lq optimal'' approximants are 
comparable with the lattice results.
}

\begin{document}

\maketitle

\section{Introduction} 
In statistical physics, the free energy density, or pressure, plays 
a central role. At high temperature, the perturbative expansion of 
the pressure of the quark-gluon plasma is known to fifth order 
\cite{arnold} in $g_s$ (the QCD coupling constant) and partially to 
sixth order \cite{kaja}. These results are reliable only when $g_s$ 
is sufficiently small. Because of the asymptotic freedom of QCD, 
this is achieved at high temperature $T \gtrsim 2$ GeV which is 
order of magnitude larger than the critical temperature ($T_c$) of 
the QCD phase transition. For $T_c \lesssim T \lesssim 2$ GeV, the 
accuracy of the approximation, within the $\overline{\mbox{MS}}$ 
scheme, does not improve by the inclusion of higher order terms 
(convergence issue). Moreover, the unphysical dependence on the 
renormalization scale $\mu$ is strong. With the traditional choice 
$\mu = 2 \pi T$, the truncated perturbation series for the pressure, 
in the $\overline{\mbox{MS}}$ scheme, does not reconcile with the 
lattice results ($0.3$ GeV $\lesssim T \lesssim 0.5$ GeV). To get 
rid of these dilemmas, various kinds of resummations are proposed. 
Among those are quasi-particle models \cite{pesh}, $\Phi$-derivative 
approximations \cite{bla}, screened perturbation theory 
\cite{kar,reb}, hard thermal loop perturbation theory \cite{and}, 
optimizations based on the principles of minimal sensitivity (PMS) 
and the fastest apparent convergence (FAC) criterions \cite{reb}, 
and the Pad\'e type resummation and its variants \cite{hatsu,koe}. 

In this paper, putting aside the convergence issue, we will 
obtain\footnote{Blaizot, Iancu, and Rebhan were the first who, among 
other things, applied the PMS criterion to the perturbative pressure 
\cite{reb}. The content of this (PMS) part of Ref. 6) and our way 
of improving it will be mentioned as occasions arise later.} the 
\lq\lq optimal'' approximants for the pressure $p$ through an 
application of the PMS criterion \cite{ste}. Throughout this paper, 
for the expression $p$ (in massless QCD), we use the one 
given\footnote{How to construct it from the available calculations 
are elucidated in Ref. 9) (see also Ref. 6)).} in Ref. 9): 
\[
p = p_E + p_M + p_G \, , 
\]
where $p_E$, $p_M$, and $p_G$ come, in respective order, from the 
hard ($\sim T$), soft ($\sim g_s T$), and supersoft ($\sim g_s^2 
T$) energy regions. The scales $g_s T$ and $g^2_s T$ are the 
color-electric and color-magnetic screening scales, respectively. A 
crucial observation made in Ref. 9) is that the \lq\lq 
short-distance pressure'' $p_E$ and the \lq\lq long-distance 
pressure'' $p_{M + G}$ ($\equiv p_M + p_G$) are {\em separately 
physical quantities}, and thus both, when calculated to all orders 
of perturbation theory, are independent of the renormalization 
scheme. Then we will apply the PMS criterion to $p_E$ and 
$p_{M + G}$ separately\footnote{Since little information is 
available on $p_G$, we will apply the PMS criterion to $p_{M + G}$ 
and not separately to $p_M$ and $p_G$. Incidentally, in Ref. 6), the 
PMS criterion is applied to the total pressure $p = p_E + p_{M + 
G}$.}. 

In \S2, a brief review of the PMS criterion is given. Then, we 
present the perturbation expansion of $p_E$ and $p_{M + G}$ in the 
form given in Ref. 9), which is taken as the basis of our 
analysis. In \S3, we apply the PMS criterion to $p_E$ and $p_{M 
+ G}$ separately and find the optimal approximants. In \S4, we 
compare our results with the lattice results and the 
$\overline{\mbox{MS}}$ results. \S5 is devoted to discussion and 
conclusion. 
\section{Preliminaries} 
\subsection{Brief review of the Principles of Minimal Sensitivity 
(PMS) (Ref. 10))}
The couplant $\alpha_s = g_s^2 / (4 \pi)$ in massless QCD runs 
according to the renormalization group equation, 
\begin{equation}
\mu \frac{\partial \alpha_s}{\partial \mu} = \beta (\alpha_s) 
= - \alpha_s^2 \sum_{j = 0}^{\infty} \beta_j \alpha_s^j \, . 
\label{RGE} 
\end{equation}
Here $\mu$ is the renormalization scale and 
\begin{eqnarray}
\beta_0 = \frac{1}{2 \pi} \left( 11 - \frac{2}{3} \, n_f \right) 
\, , \nonumber \\ 
\beta_1 = \frac{1}{4 \pi^2} \left( 51 - \frac{19}{3} \, 
n_f \right) \, , 
\label{beta01} 
\end{eqnarray}
with $n_f$ is number of quark flavors. The renormalization schemes 
can be labelled by a countable infinity of parameters ($\mu$, 
$\beta_2$, $\beta_3$, ...). [$\beta_0$ and $\beta_1$ are independent 
of the renormalization scheme.] In the $\overline{\mbox{MS}}$ 
scheme, 
\begin{equation}
(\beta_2)_{\overline{\mbox{\scriptsize{MS}}}} = \frac{1}{64 \pi^3} 
\left( 2857 - \frac{5033}{9} \, n_f + \frac{325}{27} \, n_f^2 
\right) \, . 
\label{beta2} 
\end{equation}
The dependence of $\alpha_s$ on $\beta_j$ ($j = 2, 3, ...$) 
is governed by 
\begin{eqnarray}
\frac{\partial \alpha_s}{\partial \beta_j} & = & - \beta (\alpha_s) 
\int_0^{\alpha_s} \frac{x^{j + 2}}{\left( \beta (x) \right)^2} \, d 
x \;\; \left( \equiv \gamma_j (\alpha_s) \right) \, , 
\label{RGE1} \\ 
&=& \frac{2 \pi}{9 ( j - 1)} \alpha_s^{j + 1} + ... \;\;\;\;\;\;\;\;
\;\; (j = 2, 3, ...) \, . 
\label{2d} 
\end{eqnarray}
It is to be noted that a change in $\mu$ results in a change in 
$\alpha_s$ of $O (\alpha_s^2)$ while a change in $\beta_i$ $(i = 2, 
3, ...)$ results in a change in $\alpha_s$ of $O (\alpha_s^{i + 
1})$. The running $\alpha_s = \alpha_s (\mu, \beta_2, \beta_3, ...)$ 
is obtained by solving Eqs. (\ref{RGE}) and (\ref{RGE1}) under the 
boundary condition that is provided by the experiment. 

Let ${\cal R}$ be some physical quantity, ${\cal R} = \alpha_s^l 
\sum_{j = 0}^\infty r_j \alpha_s^j$. Then, when calculated to all 
orders of perturbation theory, ${\cal R}$ is independent of the 
renormalization scheme, 
or, equivalently, of the parameters $\mu$, $\beta_2$, $\beta_3$, 
... . 

The $i$-th order perturbative approximation ${\cal R}^{(i)}$ of 
${\cal R}$ is defined by (i) truncating the perturbation series for 
${\cal R}$ at $i$th order, and (ii) replacing the expansion 
parameter $\alpha_s$ by its $i$th-order approximant defined as the 
solution to Eqs. (\ref{RGE}) and (\ref{RGE1}) with $\beta$ on the 
right-hand sides (RHS's) truncated at $i$th order; 
\begin{eqnarray} 
{\cal R}^{(i)} & = & \alpha_s^l  \sum_{j = 0}^{i - 1} r_j \alpha_s^j 
\, , \;\;\;\;\;\;\;\;\; \beta^{(i)} (\alpha_s) = - \alpha_s^2 
\sum_{j = 0}^{i - 1} \beta_j \alpha_s^j \, , 
\label{tra1} \\ 
\mu \frac{\partial \alpha_s}{\partial \mu} &=& \beta^{(i)} 
(\alpha_s) \, , \;\;\;\;\;\;\;\;\ \frac{\partial \alpha_s}{\partial 
\beta_j} = \gamma_j^{(i)} (\alpha_s) \, , 
\label{OPT1} 
\end{eqnarray} 
where $\gamma_j^{(i)} (\alpha_s)$ is defined by Eq. (\ref{RGE1}) 
with $\beta^{(i)}$ for $\beta$. 

The PMS criterion \cite{ste} for obtaining the optimal approximant 
for ${\cal R}$ goes as follows.
\begin{description} 
\item{1)} Compute $\beta^{(i)}$ and ${\cal R}^{(i)}$ in a particular 
renormalization scheme, say, $\overline{\mbox{MS}}$ scheme. 
\item{2)} Impose the requirement, 
\begin{equation} 
\mu \frac{\partial {\cal R}^{(i)}}{\partial \mu} = \frac{\partial 
{\cal R}^{(i)}}{\partial \beta_j} = 0 \;\;\;\;\;\;\;\;\; (j = 2, 3, 
... , i - 1) \, . 
\label{OPT2}
\end{equation} 
In the following we refer Eqs. (\ref{OPT2}) and (\ref{OPT1}) to as 
the OPT equations. The OPT equations determine first (a) the 
dependence of the coefficients $(r_1, ... , r_{i - 1})$ in Eq. 
(\ref{tra1}) on $(\mu, \beta_2, ... , \beta_{i - 1})$, and then (b) 
the values for the parameters $(\mu, \beta_2, ... , \beta_{i - 1}) = 
(\mu^{(i)}, \beta_2^{(i)}, ... , \beta_{i - 1}^{(i)})$ together with 
$\alpha_s = \alpha_s^{(i)}$. From (a) and (b), the values for the 
coefficients are determined, $(r_1, ... , r_{i - 1}) = (r_1^{(i)}, 
... , r_{i - 1}^{(i)})$. 
\item{3)} The optimal approximant for ${\cal R}$ is obtained by 
substituting thus determined $\alpha_s^{(i)}$ and $(r_1^{(i)}, ... , 
r_{i - 1}^{(i)})$ for, in respective order, $\alpha_s$ and $(r_1, 
... , r_{i - 1})$ in ${\cal R}^{(i)}$, Eq. (\ref{tra1}). 
\end{description} 

It should be noted here that the above definition of ${\cal 
R}^{(i)}$ is not unique. Instead of (i) of the definition, one can 
adopt 
\begin{equation} 
{\cal R}^{(i)} = \alpha_s^l \left( \sum_{j = 0}^{i - 1} r_j 
\alpha_s^j + \sum_{j = i}^\infty r_j \alpha_s^j \right) . 
\label{6} 
\end{equation} 
Here the coefficients $(r_i, r_{i + 1}, ...)$ are arbitrarily 
chosen fixed ones within the \lq\lq reasonable'' range. The 
original PMS is the special case, $r_i = r_{i + 1} = ... = 0$, of 
this more general setup. The PMS criterion can also be applied to 
more general case, e.g., 
\begin{equation}
{\cal R} = \sum_{j = 0}^\infty \alpha_s^{l + j} \sum_{k = 0}^{n_j} 
\left( r_{jk} + \sqrt{\alpha_s} r_{jk}' \right) \ln^k \alpha_s \, , 
\label{6d}
\end{equation}
where $n_j$ is $0$ or some integer. 
\subsection{Quark-gluon plasma pressure \cite{koe}} 
As has been mentioned in \S1, the form for the pressure $p$ given 
in Ref. 9) consists of three pieces, $p = p_E + p_M + p_G$. We 
will apply the PMS criterion to $p_E$ and $p_{M + G}$ $(\equiv p_M + 
p_G)$ separately. To define the border between the soft and hard 
energy regions, the scale parameter $\Lambda_E$ is introduced, $O ( 
g_s T) < O (\Lambda_E) < O (T)$. 

The long-distance pressure $p_{M + G}$ is given \cite{koe} as an 
expansion in powers of the effective EQCD parameters $g_E$ and 
$m_E$: 
\begin{eqnarray}
p_{M + G} &=& \frac{2}{3 \pi} T m_E^3 \tilde{p}_{M + G} \, , 
\label{yawa0} \\ 
\tilde{p}_{M + G} &=& 1 - \frac{9}{4 \pi} \frac{g_E^2}{m_E} \left( 
\frac{3}{4} + \ln \frac{\Lambda_E}{2 m_E} \right) - \frac{110.2}{(4 
\pi)^2} \left( \frac{g_E^2}{m_E} \right)^2 \nonumber \\ 
&& + \frac{81}{(4 \pi)^3} \left( \frac{g_E^2}{m_E} \right)^3 \left[ 
4.44 \ln \frac{\Lambda_E}{2 m_E} + 1.566 \ln 
\frac{\Lambda_E}{6 g_E^2} - 1.392 + \delta_G \right] \nonumber \\ 
&& - \frac{5}{4 \pi} (9 - n_f) \frac{\alpha_s^2}{m_E} T \, , 
\label{yawa} 
\end{eqnarray}
where 
\begin{eqnarray}
m_E^2 &=& 4 \pi (1 + n_f / 6 ) \alpha_s T^2 \left[ 
1 + \alpha_s \left( \beta_0  \ln \frac{\mu}{2 \pi T} 
+ \frac{1}{\pi} P_m (n_f) \right) \right] \, , 
\label{debye} \\ 
g_E^2 &=& 4 \pi \alpha_s T \left[ 1 + \alpha_s \left( 
\beta_0 \ln \frac{\mu}{2 \pi T} + \frac{1}{\pi} P_g (n_f) \right) 
\right] \, , 
\label{eq} 
\end{eqnarray}
with 
\begin{eqnarray}
P_m (n_f) &=& \frac{0.6124 - 0.4881 n_f - 0.04280 n_f^2}{1 + n_f / 
6} \, , \nonumber \\ 
P_g (n_f) &=& - 0.3876 - 0.4235 n_f \, . 
\end{eqnarray}
The dimensionless number $\delta_G$ in Eq. (\ref{yawa}) is unknown. 
We will allow, as in Ref. 9), the following variation: 
\begin{equation} 
- 5 < \delta_G < + 5 \, . 
\label{11de} 
\end{equation} 
We refer \cite{reb} the expression (\ref{yawa0}) - (\ref{yawa}) to 
as the untruncated form for $p_{M + G}$. 

Expanding $p_{M + G}$, Eqs. (\ref{yawa0}) and (\ref{yawa}), in 
powers of $\alpha_s^{1 / 2}$ and keeping up to and including $O 
(\alpha_s^3)$ contribution, we obtain the truncated form for 
$p_{M + G}$, 
\begin{eqnarray} 
p_{M + G}^{(2)} &=& p_{1, \, M + G}^{(2)} + p_{2, \, M + G}^{(2)} 
\, , 
\label{bunka} \\ 
p_{1, \, M + G}^{(2)} &=& \alpha_s^{3/2} \left( r_{1 0} + r_{1 1} 
\alpha_s \right) \, , 
\label{bunka3} \\ 
p_{2, \, M + G}^{(2)} &=& \alpha_s^2 \left( r_{2 0} + r_{2 1} 
\alpha_s \right) \, . 
\label{bunkai} 
\end{eqnarray} 
We see from these expressions that, for $p_{M + G}$, $\mu$ is the 
only renormalization-scheme parameter and $\beta_2$ is not. The 
reason is as follows. On 
the one hand the highest-order term in $p_{M + G}^{(2)}$ is of $O 
(\alpha_s^3)$ (see Eq. (\ref{bunkai})). On the other hand, a change 
in $\beta_2$ results in a change in $p_{M + G}^{(2)}$ of $O 
(\alpha_s^{7 / 2})$ $(< O (\alpha_s^3))$ (cf. Eq. (\ref{bunka3}) and the 
comment above after Eq. (\ref{2d})). Thus, $\beta_2$ is not the 
renormalization-scheme parameter. When the $O (\alpha_s^{7 / 2})$ 
term of $p_{M + G}$ is computed, $\beta_2$ becomes a 
renormalization-scheme parameter. 

The short-distance pressure $p_E$ is given \cite{koe} as an 
expansion in powers of the QCD couplant $\alpha_s$: 
\begin{eqnarray}
p_E &=& p_{\mbox{\scriptsize{SB}}} \left( 1 - \frac{15 (1 + 5 n_f / 
12)}{4 (1 + 21 n_f / 32)} \tilde{p}_E \right) \, , 
\label{kata0} \\  
p_{\mbox{\scriptsize{SB}}} &=& \frac{8 \pi^2}{45} \left( 1 + 
\frac{21}{32} \, n_f \right) T^4 \, , 
\label{SB} \\ 
\tilde{p}_E &=& \frac{\alpha_s}{\pi} \left[ 1 + \frac{\alpha_s}{\pi} 
\left( \pi \beta_0 \ln \frac{\mu}{2 \pi T} - 36 \frac{1 + n_f / 
6}{1 + 5 n_f / 12} \left( \ln \frac{\Lambda_E}{2 \pi T} + P_h (n_f) 
\right) \right) \right. \nonumber \\ 
&& + \left( \frac{\alpha_s}{\pi} \right)^2 \left( (\pi \beta_0)^2 
\ln^2 \frac{\mu}{2 \pi T} + F (\Lambda_E/ (2 \pi T)) \right. 
\nonumber \\ 
&& \left. \left. + \left\{ \pi^2 \beta_1 - 72 \pi \beta_0 \frac{1 + 
n_f / 6}{1 + 5 n_f / 12} \left( \ln \frac{\Lambda_E}{2 \pi T} + 
P_h (n_f) \right) \right\} \ln \frac{\mu}{2 \pi T} + \delta_E 
\right) \right] \, , \nonumber \\ 
&& 
\label{kata}
\end{eqnarray}
where 
\begin{eqnarray}
P_h (n_f) & = & \frac{244.9 + 17.24 n_f - 0.4150 
n_f^2}{135 (1 + n_f / 6)} \, , \nonumber \\ 
F (\Lambda_E/ 2 \pi T) & = & \frac{478.4 + 35.14 n_f + 4.081 
n_f^2}{1 + 5 n_f / 12} \left( \ln \frac{\Lambda_E}{2 
\pi T} + P_h (n_f) \right) \, . 
\label{maae}
\end{eqnarray}
The renormalization-scheme parameters for $p_E$ is $\mu$ and 
$\beta_2$. Eq. 
(\ref{kata}) includes the unknown number $\delta_E$, which depends 
on the parameter $\beta_2$. To see this 
dependence, we use the 
fact that the physical quantity $p_E$, Eq. (\ref{kata0}), is 
independent of the parameters, $\mu$ and 
$\beta_2$, up to and 
including $O (\alpha^3_s)$. Then from Eqs. (\ref{2d}) and 
(\ref{kata}), we find the relation between $\delta_E$ in the 
renormalization scheme with 
$\beta_2$ and its $\overline{\mbox{MS}}$ scheme counterpart $\left( 
\delta_E \right)_{\overline{\mbox{\scriptsize{MS}}}}$: 
\begin{equation}
\delta_E = \left( \delta_E 
\right)_{\overline{\mbox{\scriptsize{MS}}}} - 
\frac{\pi^2}{\beta_0} \left[ \beta_2 - 
\left( \beta_2 \right)_{\overline{\mbox{\scriptsize{MS}}}} \right] 
\, . 
\label{jyuu}
\end{equation}
For the unknown $\left( \delta_E 
\right)_{\overline{\mbox{\scriptsize{MS}}}}$, 
following Ref. 9), we will allow the following variation: 
\begin{equation} 
- | F (\Lambda_E / (2 \pi T)) | < \left( \delta_E 
\right)_{\overline{\mbox{\scriptsize{MS}}}} < + | F (\Lambda_E / 
(2 \pi T)) | \, . 
\label{11d} 
\end{equation} 
While the pure number $\left( \delta_E 
\right)_{\overline{\mbox{\scriptsize{MS}}}}$ is independent of any 
scale, $F$ will have only slight dependence on temperature $T$. 

Finally, we note that, when the expression $p_{M + G}^{(2)}$, Eq. 
(\ref{bunka}), is used for the long-distance pressure, the total 
pressure $p_{M + G}^{(2)} + p_E$ is independent \cite{koe} of 
the \lq\lq separation scale'' $\Lambda_E$ as it should be. It is 
worth remarking in passing that, if $\mu$ in $p_{M + G}^{(2)}$ is 
not equal to $\mu$ in $p_E$, this is not the case. 
\section{Optimum approximants}
It is found in Ref. 6) that various approximants, which include 
the optimal approximant, constructed using the untruncated form 
(\ref{yawa0}) - (\ref{yawa}) for $p_{M + G}$ are more reliable than 
those constructed using the truncated form 
(\ref{bunka}) - (\ref{bunkai}). It is found in Ref. 9) that this 
is also the case for the approximants constructed on the basis of 
the Pad\'e and Borel-Pad\'e resummations of $p_{M + G}$. Keeping 
these observations and the remarks made above at the end of \S2.1, 
we will also take more \lq\lq reliable form'' (\ref{yawa0}) - 
(\ref{yawa}) for $p_{M + G}$, and apply the PMS criterion to it. 

Before entering into our analysis, it is worth summarizing here the 
procedure and the results of applying the PMS criterion in Ref. 6). 
\begin{enumerate} 
\item The PMS criterion is applied to the total pressure $p = p_E + 
p_{M + G}$. 
\item The $2$-loop running couplant $\alpha_s^{(2)} (\mu)$ is used, 
so that the renormalization-scheme parameter is $\mu$ only.  
\item $\Lambda_E$ is chosen to be $\Lambda_E = \mu$. 
\end{enumerate} 
The resultant optimal approximants are close to the lattice results 
for $T \gtrsim 0.7$ GeV. The three-loop optimal pressure (with three 
quark flavors) is $\sim 30 \%$ larger than the lattice result at $T 
\simeq 0.3$ GeV, and the four-loop optimal pressure in pure-glue QCD 
with appropriate choice for $\delta_E$ and $\delta_{M + G}$ is 
$\sim 20 \%$ larger at $T \simeq 0.5$ GeV. 

We will improve 1., 2., and 3. above as follows: 
\begin{description} 
\item{1'.} The PMS criterion is applied to the short-distance 
pressure $p_E$ and the long-distance pressure $p_{M + G}$ 
separately. 
\item{2'.} For $p_E$, the three-loop running couplant 
$\alpha_s^{(3)} (\mu)$ is used\footnote{As observed above in 
conjunction with Eq. (\ref{jyuu}), when the PMS criterion is applied 
to $p_E$ (Eqs. (\ref{kata0}) and (\ref{kata})) in a consistent 
manner, the $3$-loop running couplant should be used.}, while, for 
$p_{M + G}$, the two-loop $\alpha_s^{(2)} (\mu)$ is used (see above 
after Eq. (\ref{bunkai})). 
\item{3'.} $\Lambda_E$ is determined by requiring the minimal 
sensitivity of $p$ on $\Lambda_E$ (cf. Eq. (\ref{ts}) below). 
\end{description} 

In the following,, the \lq\lq long-distance'' (\lq\lq 
short-distance'') $\mu$ and $\alpha_s$ are denoted by $\mu_{M+G}$ 
and $\alpha_s^{M + G}$ ($\mu_E$ and $\alpha_s^E$), respectively. 
The OPT equations for $p_{M + G}$ read (see above after Eq. 
(\ref{bunkai}))
\begin{equation} 
\mu_{M + G} \frac{\partial p_{M + G}}{\partial \mu_{M + G}} = 0 \, , 
\;\;\;\;\;\;\; \mu_{M + G} \frac{\partial \alpha_s^{M + G}}{\partial 
\mu_{M + G}} = \beta^{(2)} (\alpha_s^{M + G}) \, , 
\label{soft} 
\end{equation} 
where $p_{M + G}$ is as in Eq. (\ref{yawa0}) with Eq. (\ref{yawa}). 
The OPT equations for $p_E$ read 
\begin{eqnarray}
\mu_E \frac{\partial p_E}{\partial \mu_E} &=& 0 \, , \;\;\;\;\;\;\; 
\frac{\partial p_E}{\partial \beta_2} = 0 \, , 
\label{gene1} \\ 
\mu_E \frac{\partial \alpha_s^E}{\partial \mu_E} &=& \beta^{(3)} 
(\alpha_s^E) \, , \;\;\;\;\;\;\;\; 
\frac{\partial \alpha_s^E}{\partial \beta_2} = \gamma^{(3)} 
(\alpha_s^E) \, , 
\label{hard}
\end{eqnarray}
where $p_E$ is as in Eq. (\ref{kata0}) with Eqs. (\ref{SB}) - 
(\ref{maae}).  

The scale parameter $\Lambda_E$ is arbitrarily chosen in the range 
$O (g_s T) < O (\Lambda_E) < O (T)$, so that $p = p_{M + G} + p_E$ 
should not depend on $\Lambda_E$. As has been mentioned at the end 
of \S2, this is the case provided that $(\alpha)$ the truncated 
expansion $p_{M + G}^{(2)}$ (in powers of $\alpha_s^{1 / 2}$) is 
used for $p_{M + G}$ and $(\beta)$ $\mu_E = \mu_{M + G}$. In our 
case, however, both $(\alpha)$ and $(\beta)$ are not fulfilled, and 
then $p$ {\em does} depend on $\Lambda_E$. To resolve this 
ambiguity, we introduce a criterion for determining $\Lambda_E$: 
Respecting the spirit of the PMS criterion, we employ the following 
determining equation for $\Lambda_E$, 
\begin{equation}
\frac{\partial p (T, \Lambda_E)}{\partial \Lambda_E} = 0 \, . 
\label{ts}
\end{equation}

Solving a set of equations (\ref{soft}) - (\ref{ts}), we obtain the 
optimized $\mu_{M + G}$ $(= \mu_L (T))$, $\alpha_s^{M + G}$ $( = 
\alpha_s^L (\mu_L (T)) )$, $\mu_E = \mu_S (T)$, $\beta_2 = 
\beta_{2 S} (T)$, $\alpha_s^E = \alpha_s^S (\mu_S (T), \beta_{2 S} 
(T))$ and $\Lambda_E (T)$ $(= \Lambda_E^{L/S} (T))$. Here \lq\lq L'' 
and \lq\lq S'' stand for \lq\lq long-distance'' and \lq\lq 
short-distance'', respectively. Using thus determined quantities 
in Eqs. (\ref{yawa0}) and (\ref{kata0}), we obtain the optimized 
pressure $p (T) = p_{M + G} (T) + p_E (T)$. Since $p = p_{M + G} + 
p_E$ is linear in $\Lambda_E$, when the $\mu_{M + G}$ , $\mu_E$ and 
$\beta_2$ are fixed to the respective optimized values, $p (\mu_{L} (T), \mu_S 
(T), \beta_{2 S} (T), \Lambda_E; T)$ is independent of $\Lambda_E$. 
\section{Results} 
We are concerned with the temperature range $0.3$ GeV $\leq T \leq 
3$ GeV. Unless otherwise stated, we will take for the number of 
active (massless) quark flavors $n_f = 3$, which is of physical 
interest. For the QCD couplant $\alpha_s (\mu)$, we take the 
reference value $\left( \alpha_s (1.2 \mbox{ GeV}) 
\right)_{\overline{\mbox{\scriptsize{MS}}}} = 0.403$. The form for 
the short-distance $\alpha_s^E (\mu_E, \beta_2)$ is obtained from 
Eq. (\ref{OPT1}), which is give in Appendix A. Following the 
procedure stated in \S3, we obtain two solutions within the 
reasonable ranges for $\mu_{M + G}$, $\mu_E$ and $\Lambda_E$. We 
refer the solution with larger (smaller) optimal approximant for $p$ 
the solution I (II). 

For the time being, we choose $\delta_G = \left( \delta_E 
\right)_{\overline{\mbox{\scriptsize{MS}}}} = 0$. Similar behaviors 
are observed for other values of $\delta_G$ and $\left( \delta_E 
\right)_{\overline{\mbox{\scriptsize{MS}}}}$. 

In Fig. 1, we present the solutions I and II against $T$. The 
results are normalized by the Stefan-Boltzmann pressure 
$p_{\mbox{\scriptsize{SB}}}$ (cf. Eq. (\ref{SB})). Also shown are 
the optimized long-distance pressure $p_{M + G}$ and the 
short-distance pressure $p_E$. The circles are the predictions of 
the lattice calculations \cite{karsch} (see Ref. 9) for details.) 
From the figures we see that i) for the solution II, $p_{M + G} < 0$ 
and ii) the solution I gives better agreement with the lattice 
result. From these observations, in the following, we adopt the 
solution I as the better candidate for the optimal approximant. 

In Fig. 2 the optimized $\mu_{M + G}$, $\mu_L (T)$, and $\mu_E$, 
$\mu_S (T)$, are shown against $T$. The dot-dashed lines shows the 
traditional choice $\mu = 2 \pi T$. The ratios $\mu_S (T)/T$ and 
$\mu_L (T)/T$ decrease monotonically with $T$: The ratio $\mu_S (T)/2 \pi T$ is
larger than 1 at $T = 0.3$ GeV, $\simeq 1$ at 
$T = 0.5$ GeV and becomes smaller than 1 for $T > 0.5$ GeV. $\mu_{L} (T)$ 
is smaller than $\mu_S (T)$ and 
$\mu_{L} (T) / \mu_S (T) \simeq  0.31\sim 0.36 $ in the range shown in 
the figure. This is in accord with an intuition: The scale $\mu_E$ 
that governs the physics in the hard region, being of $O (T)$, is 
larger than the scale $\mu_{G + M}$ in the soft region. 

Fig. 3 depicts the optimized $\beta_2$, $\beta_{2S} (T)$, against 
$T$. The dashed line shows $\beta_2 = 
(\beta_2)_{\overline{\mbox{\scriptsize{MS}}}}$. $\beta_{2S} (T)$ 
monotonically decreases with $T$ crossing the $\overline{\mbox{MS}}$ 
value at $T \simeq 0.5$ GeV. We note that Eq. (\ref{2d}) tells us 
that increase (decrease) in $\beta_2$ acts as increasing 
(decreasing) $\alpha_s$, the same effect as arising from decreasing 
(increasing) $\mu$. Then, we suspect that, if we fix $\beta_2$ to 
$\left( \beta_2 \right)_{\overline{\mbox{\scriptsize{MS}}}}$ and 
vary only $\mu$, we obtain much smaller (larger) optimized $\mu_E$ 
for $T \lesssim 0.5$ ($T \gtrsim 0.5$) GeV. Thus, we expect that the 
optimized $\mu_E$ tends to $2 \pi T$. It is amazing to note that, at 
$T \sim 0.5$ GeV, $\mu_S (T)$ and $\beta_{2S} (T)$ are close to 
the traditional choice  $\mu =2 \pi T$ and the $\overline{\mbox{MS}}$ value, respectively. 

In Fig. 4, we show the optimized $\Lambda_E$, $\Lambda_E^{L/S} (T)$, 
and $m_E (T)$ against $T$. For reference $\mu_L (T)$ in Fig. 2 is also shown again. 
We see that 
$\Lambda_E^{L/S} (T) \sim  m_E (T) \sim \mu_L (T) \: (< \mu_{S}(T))$. 
This is 
not in accord with an intuitive expectation $\mu_{M + G} < \Lambda_E 
< \mu_E$. However, as seen at the end of \S3, when $\mu_{M + G}$, 
$\mu_E$, and $\beta_2$ are fixed to their respective optimized 
values, the pressure $p$ does not depend on $\Lambda_E$. 

In Fig. 5, we show the pressure $p$ at $T = 0.5$ GeV near the 
optimized values of $\alpha_s^E \simeq \alpha_s^S (T)$, $\alpha_s^{M + 
G} \simeq \alpha_s^L (T)$, $\beta_2 \simeq \beta_{2 S} (T)$ and 
$\Lambda_E \simeq \Lambda_E^{L/S} (T)$. In Fig. 5 (a), we show the 
behavior of $p$ against $\alpha_s^{S}$, when other quantities, 
$\alpha_s^{M + G}$, $\beta_2$ and $\Lambda_E$ are fixed to the 
respective optimized values. The arrow indicates the optimized value of 
$\alpha_s^E $. Similarly, Figs. 5 (b) and (c) show the 
dependencies of $p$ on $\alpha_s^{M + G}$ and on $\beta_2$, 
respectively. From Fig. 5(c), we see that $p$ depends on $\beta_2$ 
only weakly. As has been repeatedly pointed out, 
$p$ does not depend on $\Lambda_E$, when $\alpha_s^E$, $\alpha_s^{M+G}$ 
and $\beta_2$ are fixed to the respective optimized values. 

In Fig. 6, we present the optimal approximants for p, when the unknown 
constants $\left( \delta_E 
\right)_{\overline{\mbox{\scriptsize{MS}}}}$ and $\delta_{G}$ are 
varied. Within the ranges, (\ref{11de}) and (\ref{11d}), of the 
unknown constants, $(\delta_G, \left( \delta_E 
\right)_{\overline{\mbox{\scriptsize{MS}}}}) = (+ 5, - F)$ yields 
the maximum $p / p_{\mbox{\scriptsize{SB}}}$, while $(\delta_G, 
\left( \delta_E \right)_{\overline{\mbox{\scriptsize{MS}}}}) = 
(- 5, + F)$ yields the minimum $p / p_{\mbox{\scriptsize{SB}}}$. 
For comparison, the $\overline{\mbox{MS}}$ results are also 
shown\footnote{$p_{M + G}$ (Eq. (\ref{yawa})) contains the 
logarithmic factors $\ln m_E$ and $\ln g_E^2$, for which we have 
expanded as $\ln \alpha_s + c_1 + c_2 \alpha_s + ...$ with 
respective constants $c_1$, $c_2$ , ... .  Then, for $p_{M + G}$, we 
have used the truncated forms (\ref{bunka}) - (\ref{bunkai}) and 
then $p$ $(= p_{M + G} + p_E$) is independent of $\Lambda_E$. For 
the $\beta (\alpha_s)$ function, we have used the three-loop form. 
For the scale parameter $\mu$, we have used the conventional one $\mu = 2 
\pi T$. The unknown constant $\left( \delta_E 
\right)_{\overline{\mbox{\scriptsize{MS}}}}$ in Eq. (\ref{kata}), we 
vary in the range (\ref{11d}). For $\Lambda_E$ in $F (\Lambda_E / 
2 \pi T)$ there, we have used Eq. (\ref{maae}) with $\Lambda_E = 
\sqrt{2 \pi T m_E (T)}$. For $\mu_{M + G}$ and $\alpha_s^{M + G}$ in 
$m_E (\mu_{M + G}, \alpha_s^{M + G}; T)$, we have substituted 
the respective optimized values.}. From the figure, we see as a 
bonus that the resultant optimal approximants depend rather weakly 
on $\delta_G$ and $\left( \delta_E 
\right)_{\overline{\mbox{\scriptsize{{MS}}}}}$. Our results are 
substantially improved when compared to the $\overline{\mbox{MS}}$ 
results. Taking into account that the lattice results should be 
taken within an error of $10-15$ \%, our results are well 
compatible with the lattice results. 

Finally, in Fig. 7, we depicts the optimal approximants (solutions 
I) for the case of pure glue QCD ($n_f = 0$). For comparison, the 
solution II for $\delta_G = \left( \delta_E 
\right)_{\overline{\mbox{\scriptsize{MS}}}} = 0$ is also shown. 
Same comments as for the $n_f = 3$ case applies here. For 
$\delta_G = \left( \delta_E \right)_{\overline{\mbox{\scriptsize{MS}}}} = 0$, 
the $\overline{\mbox{MS}}$ result accidentally coincides with the 
solution II with high precision. Although the 
results are substantially improved when compared to the 
$\overline{\mbox{MS}}$ results (except for the case of $(\delta_G, \left( \delta_E 
\right)_{\overline{\mbox{\scriptsize{MS}}}}) = (+ 5, - F)$), 
agreement with the lattice results 
are less than that for the realistic case of $n_f = 3$. 
\section{Discussion and Conclusion} 
The QCD pressure $p$ has been computed within the improved 
perturbation theory using the $\overline{\mbox{MS}}$ renormalization 
scheme, up to and including the $O (g^6_s \ln (1 / g_s))$ and part 
of the $O (g^6_s)$ terms. The truncated perturbation series for $p$ 
does not reconcile with the relatively low $T$ lattice results. 
As has been mentioned in \S1, various approaches are proposed 
for improving the perturbative pressure $p$. Each approach 
performs a sort of resummation on the basis of respective 
philosophy. In this paper we add one more to them. 

We have applied the Principles of Minimal Sensitivity (PMS) 
criterion to $p$. This is not new, but in Ref. 6), among other 
things, the PMS criterion has already been applied to $p$. However, 
as has been mentioned in \S3, application is not made in a consistent 
manner. Meanwhile, an important observation is made in Ref. 9): 
The contributions to $p$ from the hard and soft energy regions, 
$p_E$ and $p_{M+G}$, are separately physical quantities. On the 
basis of this fact, we have applied the PMS criterion to $p_E$ and 
$p_{M+G}$ separately. For determining the factorization scale 
$\Lambda_E$, which separates hard and soft scales, we also adopt the 
minimal sensitivity criterion. 

We have found that the optimal approximants are significantly 
improved when compared to the $\overline{\mbox{MS}}$ results. 
Especially, for the realistic case of $n_f = 3$, agreement 
between our approximants and the lattice results is satisfactory. 
Moreover, we have found that, when compared to the 
$\overline{\mbox{MS}}$ results, the dependence of the approximants 
on the unknown terms of $O (g_s^6)$ is rather weak. On the other 
hand, for the case of $n_f = 0$, although the results are 
substantially improved, agreement with the lattice results are 
poorer than that for the $n_f = 3$ case. 

Further improvement may be expected if we perform the Pad\'e type 
resummation starting from the optimal approximants obtained here, 
which is outside the scope of this paper. 
\section*{Acknowledgments}
We thank A. Nakamura for instructions of lattice calculations. 
The authors thank the useful discussion at the Workshop on Thermal 
Field Theories and their Applications, held at the Yukawa Institute 
for Theoretical Physics, Kyoto, Japan, 9 - 11 August, 2004. 
A. N. is supported in part by the Grant-in-Aid for Scientific 
Research [(C)(2) No. 17540271] from the Ministry of Education, 
Culture, Sports, Science and Technology, Japan, No.(C)(2)-17540271. 
\setcounter{equation}{0}
\setcounter{section}{1}
\section*{Appendix A: Running Couplant} 
\def\theequation{\mbox{\Alph{section}.\arabic{equation}}}
The relation between the three-loop running couplants  
$\alpha_s (\mu_0, ( \beta_2 
)_{\overline{\mbox{\scriptsize{MS}}}} )$ and \\
$\alpha_s (\mu, \beta_2)$ 
is obtained by 
integrating the following two equations (cf. Eq. (\ref{OPT1})), 
\begin{eqnarray}
\mu \frac{\partial \alpha_s}{\partial \mu} & = & \beta^{(3)} 
(\alpha_s) \, , 
\label{dai1} \\ 
\frac{\partial \alpha_s}{\partial \beta_2} & = & \gamma_2^{(3)} 
(\alpha_s) \, , 
\label{dai2} 
\end{eqnarray}
along a path ${\cal C}$ in a $(\ln \mu, \beta_2)$-plane, which 
starts from the point $(\ln \mu_0, 
(\beta_2)_{\overline{\mbox{\scriptsize{MS}}}})$ and arrives at the 
point $(\ln \mu, \beta_2)$. Because of the integrability condition 
$\partial \beta^{(3)} (\alpha_s) / \partial \beta_2 = \partial 
\gamma^{(3)}_2 (\alpha_s) / \partial \ln \mu$, one can use any path 
${\cal C}$. For convenience, we choose ${\cal C} = {\cal C}_1 \oplus 
{\cal C}_2 \oplus {\cal C}_3$ with ${\cal C}_1 = (\ln \mu_0, 
(\beta_2)_{\overline{\mbox{\scriptsize{MS}}}}) \to (\ln \mu_M, 
(\beta_2)_{\overline{\mbox{\scriptsize{MS}}}})$, ${\cal C}_2 = (\ln 
\mu_M, (\beta_2)_{\overline{\mbox{\scriptsize{MS}}}}) \to (\ln 
\mu_M, \beta_2)$ and ${\cal C}_3 = (\ln \mu_M, \beta_2) \to (\ln 
\mu, \beta_2)$. At the final stage, we take the limit $\mu_M \to + 
\infty$. 

Integrating Eq. (\ref{dai1}) along the path ${\cal C}_1$, we obtain 
\begin{eqnarray}
\beta_0 \ln \left( \frac{\mu_M}{\mu_0} \right) & = & 
\frac{1}{\alpha_{sM}} - \frac{1}{\alpha_{s0}} + 
\frac{\beta_1}{\beta_0} \ln \left( \frac{\alpha_{sM}}{\alpha_{s0}} 
\right) 
- \frac{\beta_1}{2 \beta_0} \ln \left( \frac{\beta_0 + \beta_1 
\alpha_{sM} + \left( \beta_2 
\right)_{\overline{\mbox{\scriptsize{MS}}}} \alpha_{sM}^2}{\beta_0 
+ \beta_1 \alpha_{s0} + \left( \beta_2 
\right)_{\overline{\mbox{\scriptsize{MS}}}} \alpha_{s0}^2} \right) 
\nonumber \\ 
&& - \frac{\beta_1^2 - 2 \beta_0 \left( \beta_2 
\right)_{\overline{\mbox{\scriptsize{MS}}}}}{2 \beta_0^2} \left[ G 
\left( \alpha_{sM}, \left( \beta_2 
\right)_{\overline{\mbox{\scriptsize{MS}}}} \right) - G \left( 
\alpha_{s0}, \left( \beta_2 
\right)_{\overline{\mbox{\scriptsize{MS}}}} \right) \right] \, , 
\label{me}
\end{eqnarray}
where $\alpha_{s0} = \alpha_s ( \mu_0, ( \beta_2 
)_{\overline{\mbox{\scriptsize{MS}}}} )$ and $\alpha_{sM} = \alpha_s 
\left( \mu_M, \left( \beta_2 
\right)_{\overline{\mbox{\scriptsize{MS}}}} \right)$ and 
\[
G (x, y) = \frac{2 \beta_0}{\sqrt{4 \beta_0 y - \beta_1^2}} 
\arctan \left( \frac{2 x y + \beta_1}{\sqrt{4 \beta_0 y - 
\beta_1^2}} \right) \, .  
\]
Integration of Eq. (\ref{dai1}) along the path ${\cal C}_3$ yields 
a similar equation for $\beta_0 \ln ( \mu / \mu_M )$. Adding this to 
Eq. (\ref{me}), we obtain the formula for $\beta_0 \ln (\mu / 
\mu_0)$, which includes, among others, $\alpha_s \left( \mu_M, 
\left( \beta_2 \right)_{\overline{\mbox{\scriptsize{MS}}}} \right)$ 
and $\alpha_s (\mu_M, \beta_2)$. From Eq. (\ref{dai2}), we see that 
$\alpha_s (\mu_M, (\beta_2 )_{\overline{\mbox{\scriptsize{MS}}}}) = 
\alpha_s (\mu_M, \beta_2 ) + O (\alpha_s^3)$. Substituting this for 
$\alpha_s (\mu_M, (\beta_2 )_{\overline{\mbox{\scriptsize{MS}}}})$ 
in the formula obtained above and taking the limit $\mu_M \to + 
\infty$ ($\alpha_s (\mu_M, \beta_2) \to 0$), we see that the $O 
(\alpha_s^3)$ term does not survive. 

Thus, we finally obtain 
\begin{eqnarray}
\beta_0 \ln \left( \frac{\mu}{\mu_0} \right) & = & 
\frac{1}{\alpha_s} - \frac{1}{\alpha_{s0}} + \frac{\beta_1}{\beta_0} 
\ln \left( \frac{\alpha_s}{\alpha_{s0}} \right) - \frac{\beta_1}{2 
\beta_0} \ln \left( \frac{\beta_0 + \beta_1 \alpha_s + \beta_2 
\alpha_s^2}{\beta_0 + \beta_1 \alpha_{s0} + \left( \beta_2 
\right)_{\overline{\mbox{\scriptsize{MS}}}} \alpha_{s0}^2} \right) 
\nonumber \\ 
&& - \frac{\beta_1^2 - 2 \beta_0 \left( \beta_2 
\right)_{\overline{\mbox{\scriptsize{MS}}}}}{2 \beta_0^2} \left[ 
G \left( 0, \left( \beta_2 
\right)_{\overline{\mbox{\scriptsize{MS}}}} \right) - G \left( 
\alpha_{s0}, \left( \beta_2 
\right)_{\overline{\mbox{\scriptsize{MS}}}} \right) \right] 
\nonumber \\ 
&& - \frac{\beta_1^2 - 2 \beta_0 \beta_2}{2 \beta_0^2} [ G 
(\alpha_s, \beta_2) - G (0, \beta_2)] \, , 
\label{fine}
\end{eqnarray}
where $\alpha_s = \alpha_s (\mu, \beta_2)$. 

\begin{figure}[h]
\begin{center}
\includegraphics[width=15cm,clip]{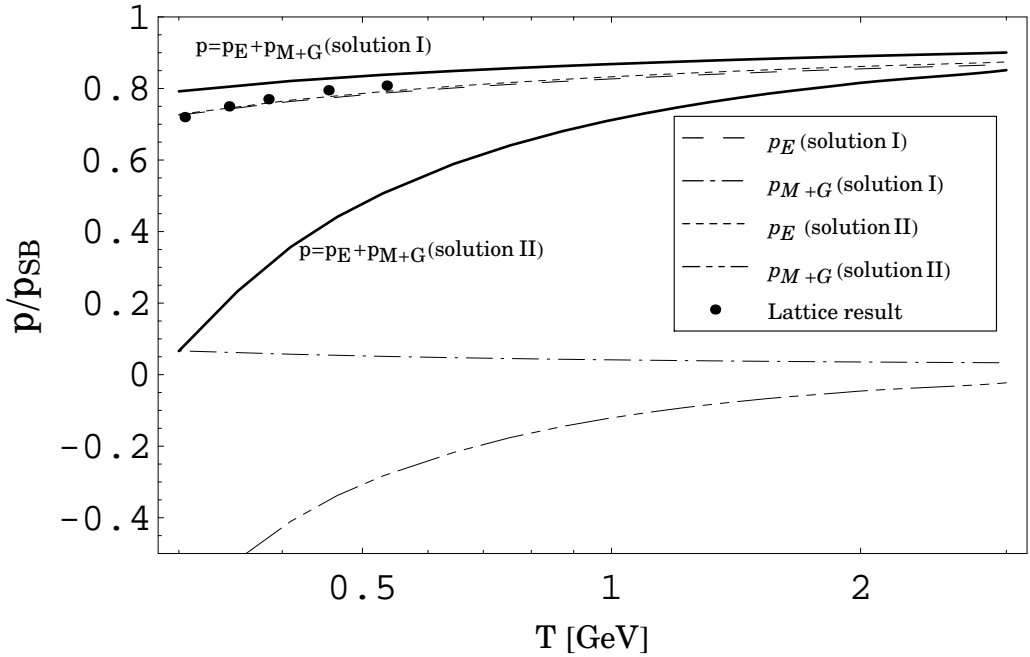}
\end{center}
\caption{The solutions I and II for the optimal approximant, 
$p / p_{\mbox{\scriptsize{SB}}}$, as functions of $T$, when 
$\delta_G = \left( \delta_E 
\right)_{\overline{\mbox{\scriptsize{MS}}}} = 0$. 
($p_{\mbox{\scriptsize{SB}}}$ is the Stefan-Boltzmann pressure.) 
The optimized long-distance pressure $p_{M + G}$ and the 
short-distance pressure $p_E$ are also separately shown. The 
circles are the predictions of the lattice calculations.} 
\end{figure}

\begin{figure}[h]
\begin{center}
\includegraphics[width=10cm,clip]{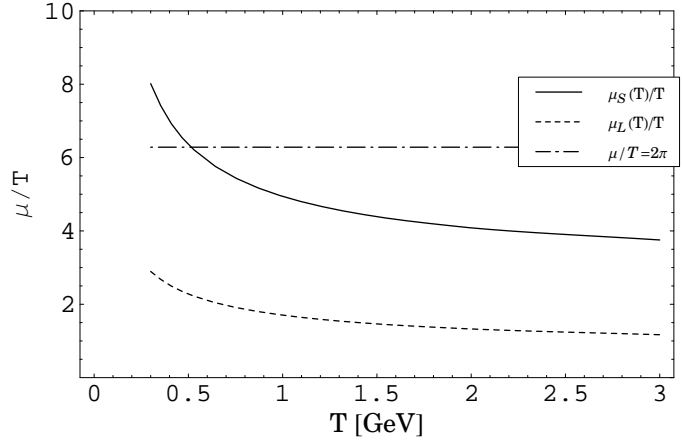}
\end{center}
\caption{
The optimized $\mu_{M + G}$, $\mu_L (T)$, and $\mu_E$, $\mu_S (T)$, 
for $\delta_G = \left( \delta_E 
\right)_{\overline{\mbox{\scriptsize{MS}}}} = 0$ as functions of 
$T$. The dot-dashed lines shows the traditional choice $\mu = 2 \pi 
T$.}
\end{figure}

\begin{figure}[h]
\begin{center}
\includegraphics[width=9cm,clip]{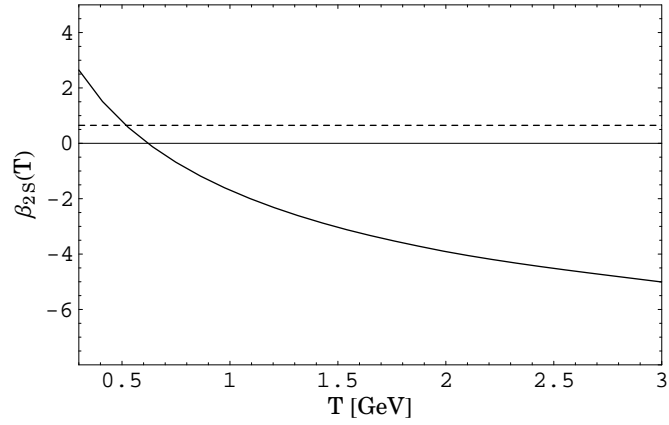}
\end{center}
\caption{
The optimized $\beta_2$, $\beta_{2S} (T)$, as a function of $T$. The dashed line 
shows $\beta_2 = (\beta_2)_{\overline{\mbox{\scriptsize{MS}}}}$.} 
\end{figure}

\begin{figure}[h]
\begin{center}
\includegraphics[width=10cm,clip]{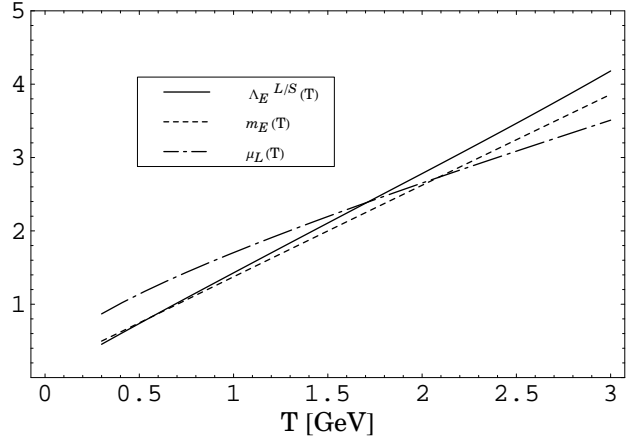}
\end{center}
\caption{The optimized $\Lambda_E$, $\Lambda_E^{L/S} (T)$, and $m_E$ 
as functions of $T$. For reference, the optimized $\mu_{M + G}$, 
$\mu_L (T)$, is also shown.} 
\end{figure}

\begin{figure}[h]
\begin{center}
\begin{minipage}{.30\linewidth} 
  \includegraphics[width=6cm,clip]{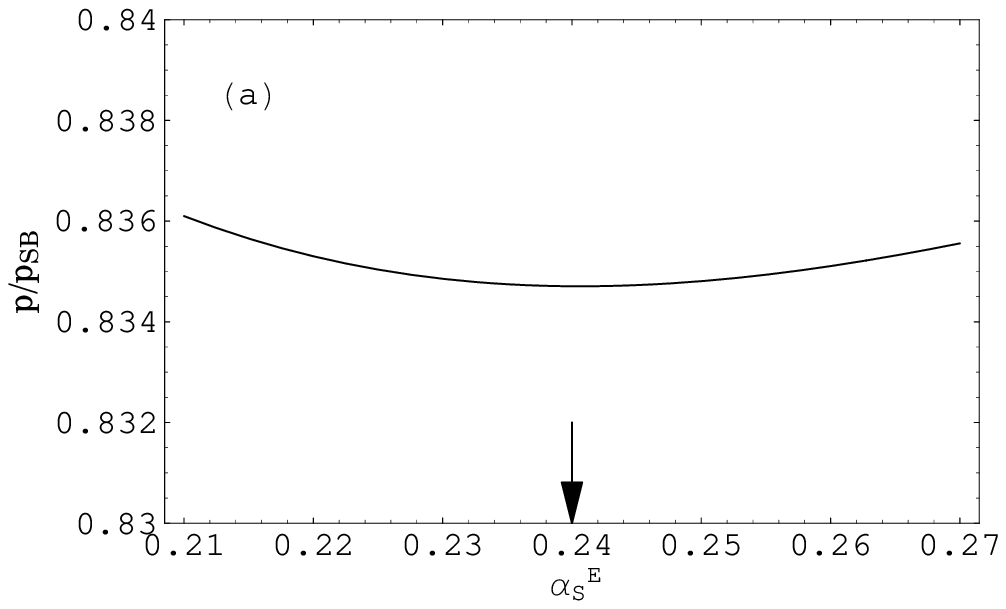} 
  \end{minipage}
\hspace{5.0pc} 
\begin{minipage}{.30\linewidth} 
  \includegraphics[width=6cm,clip]{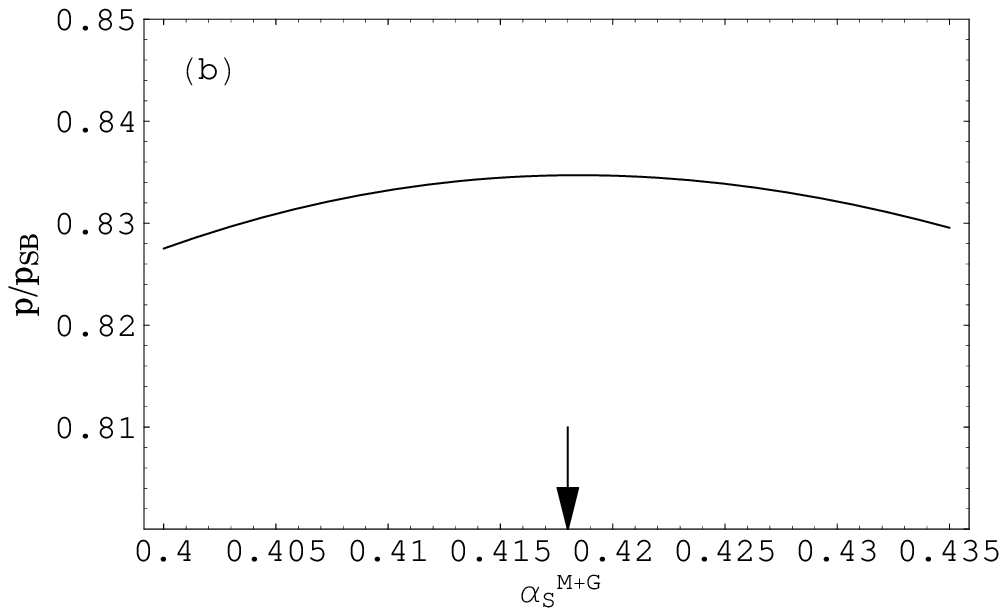} 
  \end{minipage}
\end{center} 
\begin{center}
\begin{minipage}{.30\linewidth} 
  \includegraphics[width=6cm,clip]{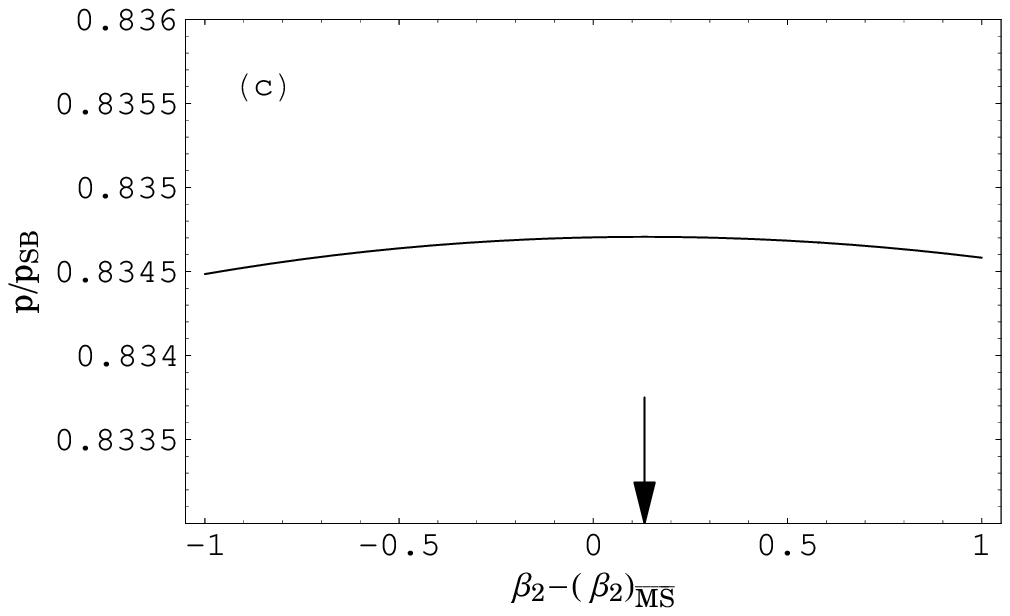} 
  \end{minipage}
\end{center}
\caption{
The pressure $p$ near the optimized values of $\alpha_s^E$, 
$\alpha_s^{M + G}$ and $\beta_2$, when $T = 0.5$ Gev. The arrows 
in Figs. (a), (b) and (c) indicate the optimized values of $\alpha_s^E$, 
$\alpha_s^{M + G}$ and $\beta_2$, respectively. Details are given in the text. 
}
\end{figure}

\begin{figure}[h]
\begin{center}
\includegraphics[width=13.5cm,clip]{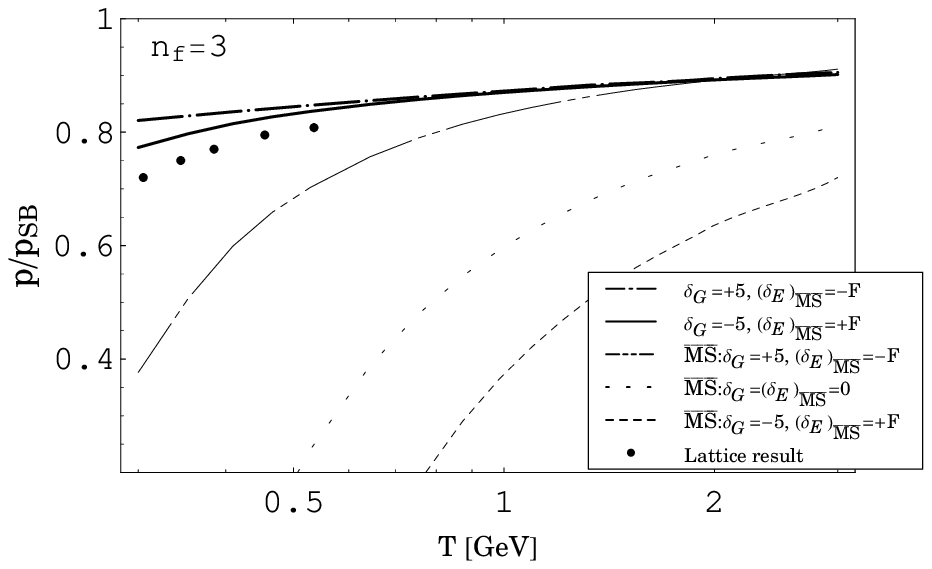}
\end{center}
\caption{
Optimal approximants (the solutions I) for $p / 
p_{\mbox{\scriptsize{SB}}}$ as functions of $T$ for various values 
of $\delta_G$ and $\left( \delta_E 
\right)_{\overline{\mbox{\scriptsize{MS}}}}$. Within the ranges, 
(\ref{11de}) and (\ref{11d}), of $(\delta_G, \left( \delta_E 
\right)_{\overline{\mbox{\scriptsize{MS}}}})$, $(\delta_G, \left( 
\delta_E \right)_{\overline{\mbox{\scriptsize{MS}}}}) = (+ 5, - F)$, 
yields the maximum $p / p_{\mbox{\scriptsize{SB}}}$ (dot-dashed 
line) while $(\delta_G, \left( \delta_E 
\right)_{\overline{\mbox{\scriptsize{MS}}}}) = (- 5, + F)$ yields 
the minimum one (solid line). For comparison, the lattice results 
and the $\overline{\mbox{MS}}$ results are also shown. Displayed 
$\overline{\mbox{MS}}$ results are the maximum one ${[}(\delta_G, 
\left( \delta_E \right)_{\overline{\mbox{\scriptsize{MS}}}}) = (+5, 
-F){]}$, within the ranges (\ref{11de}) and (\ref{11d}), the 
minimum one ${[}(\delta_G, \left( \delta_E 
\right)_{\overline{\mbox{\scriptsize{MS}}}}) = (-5, +F){]}$ and the 
one with $\delta_G = \left( \delta_E 
\right)_{\overline{\mbox{\scriptsize{MS}}}} = 0$.} 
\end{figure}

\begin{figure}[h]
\begin{center}
\includegraphics[width=17cm,clip]{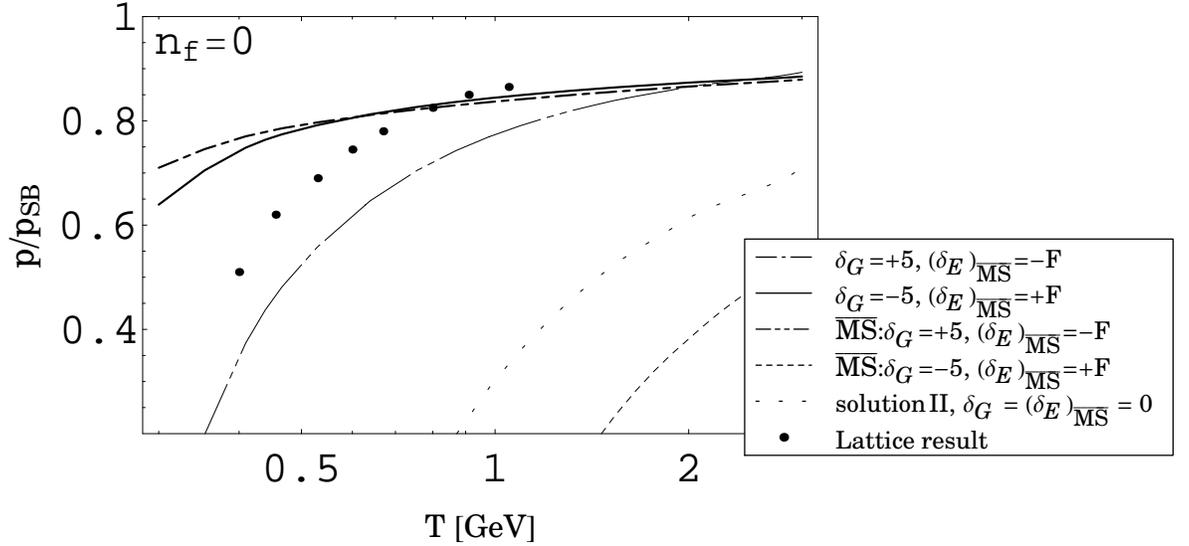}
\end{center}
\caption{
Analogous to Fig. 6 but with $n_f = 0$. The solution II for 
$\delta_G = \left( \delta_E 
\right)_{\overline{\mbox{\scriptsize{{MS}}}}} = 0$ is also shown. 
}

\end{figure}

\end{document}